\documentclass[preprint,amsmath,amssymb,11pt,english,aps,showpacs,showkeys]{revtex4}
\usepackage{graphicx}
\usepackage{graphics}
\usepackage{amsmath}
\usepackage{dcolumn}
\usepackage{amssymb}
\usepackage{bm}
\usepackage[latin1]{inputenc}

\begin{document}
\title{A geometric approach to confining a Dirac neutral particle in analogous way to a quantum dot}
\author{Knut Bakke}
\email{kbakke@fisica.ufpb.br}
\affiliation{Departamento de F\'isica, Universidade Federal da Para\'iba, Caixa Postal 5008, 58051-970, Jo\~ao Pessoa, PB, Brazil.}

\begin{abstract}
We discuss a geometric approach to confining a Dirac neutral particle with a permanent magnetic dipole moment interacting with external fields to a hard-wall confining potential in the Minkowski spacetime through noninertial effects. We discuss the behaviour of external fields induced by noninertial effects, and a case where relativistic bound states can be achieved in analogous way to having a Dirac particle confined to a quantum dot. We show that this confinement of a Dirac neutral particle analogous to a quantum dot arises from noninertial effects that give rise to the geometry of the manifold playing the role of a hard-wall confining potential. We also discuss the possible use of this mathematical model in studies of noninertial effects on condensed matter systems described by the Dirac equation.

\end{abstract}


\maketitle

%$ $ 

\section{Introduction }

Bound states for Dirac particles have been attracted the focus of several works in recent years. For instance, condensed matter systems described by the Dirac equation such as graphene \cite{graf}, topological insulators \cite{dirac2} and cold atoms \cite{dirac3}, have attracted a great deal of new studies \cite{graf1,graf2,graf3,graf4,ti1,l12,bf30}. Dirac particles have also been studied in $\mathcal{PT}$-symmetric potentials \cite{dic10}, confined to quantum potentials \cite{dic1}, and in the relativistic Landau quantization \cite{bf10} based on the Aharonov-Casher setup \cite{ac}. Dirac particles have also been studied in interferometry of dipole moments \cite{anan,sil}, by interacting with the Dirac oscillator \cite{dir3} and in the Aharonov-Bohm effect for bound states \cite{val}. It is worth mentioning the generalization of the quantum Hall effect made with Dirac particles in \cite{hall3,hall4}.

In this work, we show that noninertial effects give rise to a hard-wall confining potential that allow us to discuss a geometric approach to confining a Dirac neutral particle with a permanent magnetic dipole moment interacting with external fields in analogous way to a quantum dot. We also show that the relativistic bound states solutions analogous to having a Dirac particle confined to a quantum dot can only be achieved depending on the behaviour of external fields induced by noninertial effects. In recent years, noninertial effects have been discussed in quantum mechanics showing the appearance of phase shifts in the wave function of the quantum particle. The best famous effects are the Sagnac effect \cite{sag} and the Mashhoon effect \cite{r3}. These quantum effects have been widely discussed in recent decades \cite{sag2,sag4,r4,r5,r6}. Another well-known noninertial effect in quantum mechanics arises from the coupling between the angular momentum and the angular velocity which is called the Page-Werner {\it et al.} coupling \cite{r1,r2}. Analogue effects of the Aharonov-Casher effect \cite{ac} have been discussed in a rotating frame \cite{bf3} and in the Fermi-Walker reference frame \cite{bf14}. In the context of bound states, the noninertial effects of the Fermi-Walker reference frame have been used to achieve the Landau quantization for a neutral particle \cite{b} based on the He-McKellar-Wilkens system \cite{hmw}, and bound states for a neutral particle analogous to having a neutral particle subject to a parabolic quantum dot potential \cite{b2,b4}. The Landau quantization based on the Aharonov-Casher setup \cite{ac} has also been studied in rotating frames and in the presence of a topological defect \cite{bf16}.

The structure of this papers is: in section II, we discuss the relativistic quantum dynamics of a neutral particle with a permanent magnetic dipole moment interacting with a field configuration induced by noninertial effects, and obtain bound states for a Dirac neutral particle subject to a hard-wall confining potential which arise from noninertial effects; in section III, we present our conclusions.

\section{bound states for a Dirac neutral particle}

In this work, we consider a neutral particle with a permanent magnetic dipole moment parallel to the $z$ axis. The quantum dynamics of a neutral particle, with a permanent magnetic dipole moment interacting with external magnetic and electric fields, is described by introducing the following nonminimal coupling into the Dirac equation \cite{anan,sil,ac}
\begin{eqnarray}
i\gamma^{\kappa}\,\partial_{\kappa}\rightarrow\,i\gamma^{\kappa}\,\partial_{\kappa}+\frac{\mu}{2}\,F_{\kappa\nu}\left(x\right)\,\Sigma^{\kappa\nu},
\label{2.1}
\end{eqnarray}
where $\mu$ is the permanent magnetic dipole moment of the neutral particle, $F_{\kappa\nu}\left(x\right)$ is the electromagnetic field tensor, whose components are $F_{0i}=-F_{i0}=E_{i}$ and $F_{ij}=-F_{ji}=-\epsilon_{ijk}\,B^{k}$, and $\Sigma^{\kappa\nu}=\frac{i}{2}\left[\gamma^{\kappa},\gamma^{\nu}\right]$. Several studies of the quantum dynamics of a neutral particle based on the nonminimal coupling (\ref{2.1}) have been made in recent years \cite{anan,sil,ac}. One of the most important studies of the quantum dynamics of a neutral particle with a permanent magnetic dipole moment is called the Aharonov-Casher effect \cite{ac}. The Aharonov-Casher effect \cite{ac} has been studied in the noncommutative quantum mechanics \cite{lin2}, in the presence of topological defects \cite{bf3,bf14}, in supersymmetric quantum mechanics \cite{ac1} and in the relativistic Landau quantization \cite{bf10,bbs}. The Aharonov-Casher system \cite{ac} presents the cylindrical symmetry, thus, it is convenient to work with curvilinear coordinates \cite{schu,bf10,bbs} by using the mathematical formulation of the spinor theory in curved spacetime \cite{weinberg,bd}. In a curved spacetime background, spinors are defined locally in the reference frame of the observers \cite{weinberg,bd}. We can build a local reference frame through a noncoordinate basis $\hat{\theta}^{a}=e^{a}_{\,\,\,\mu}\left(x\right)\,dx^{\mu}$, where the components $e^{a}_{\,\,\,\mu}\left(x\right)$ satisfy the relation $g_{\mu\nu}\left(x\right)=e^{a}_{\,\,\,\mu}\left(x\right)\,e^{b}_{\,\,\,\nu}\left(x\right)\,\eta_{ab}$ \cite{bd,naka}, with $\eta_{ab}=\mathrm{diag}(- + + +)$ being the Minkowski tensor. The indices $(a,b,c=0,1,2,3)$ indicate the local reference frame of the observers. The components of the noncoordinate basis $e^{a}_{\,\,\,\mu}\left(x\right)$ are called \textit{tetrads}. The tetrads have an inverse defined as $dx^{\mu}=e^{\mu}_{\,\,\,a}\left(x\right)\,\hat{\theta}^{a}$ which satisfy the condition: $e^{a}_{\,\,\,\mu}\left(x\right)\,e^{\mu}_{\,\,\,b}\left(x\right)=\delta^{a}_{\,\,\,b}$ and $e^{\mu}_{\,\,\,a}\left(x\right)\,e^{a}_{\,\,\,\nu}\left(x\right)=\delta^{\mu}_{\,\,\,\nu}$. In this way, the $\gamma^{\mu}$ matrices given in (\ref{2.1}) correspond to that defined in curvilinear coordinates and are related to the $\gamma^{a}$ matrices (defined in the Minkowski spacetime) via $\gamma^{\mu}=e^{\mu}_{\,\,\,a}\left(x\right)\gamma^{a}$. The $\gamma^{a}$ matrices are defined in the local reference frame and correspond to the Dirac matrices in the Minkowski spacetime \cite{greiner,bd}
\begin{eqnarray}
\gamma^{0}=\hat{\beta}=\left(
\begin{array}{cc}
1 & 0 \\
0 & -1 \\
\end{array}\right);\,\,\,\,\,\,
\gamma^{i}=\hat{\beta}\,\hat{\alpha}^{i}=\left(
\begin{array}{cc}
 0 & \sigma^{i} \\
-\sigma^{i} & 0 \\
\end{array}\right);\,\,\,\,\,\,\Sigma^{i}=\left(
\begin{array}{cc}
\sigma^{i} & 0 \\
0 & \sigma^{i} \\	
\end{array}\right),
\label{2.3}
\end{eqnarray}
with $\vec{\Sigma}$ being the spin vector. The matrices $\sigma^{i}$ are the Pauli matrices and satisfy the relation $\left(\sigma^{i}\,\sigma^{j}+\sigma^{j}\,\sigma^{i}\right)=2\,\eta^{ij}$.  Now, the Dirac equation must be written by changing the partial derivative of the expression (\ref{2.1}) to the covariant derivative of a spinor \cite{bd,schu} given by $\nabla_{\mu}=\partial_{\mu}+\Gamma_{\mu}\left(x\right)$, with $\Gamma_{\mu}\left(x\right)=\frac{i}{4}\,\omega_{\mu ab}\left(x\right)\,\Sigma^{ab}$ being the spinorial connection \cite{bd,naka,schu}. Hence, the Dirac equation (in curvilinear coordinates with the interaction of the permanent magnetic dipole moment of the neutral particle with external fields) is given by the following expression (with $\hbar=c=1$)
\begin{eqnarray}
i\,\gamma^{a}\,e^{\mu}_{\,\,\,a}\left(x\right)\,\partial_{\mu}\psi+i\,\gamma^{a}\,e^{\mu}_{\,\,\,a}\left(x\right)\,\Gamma_{\mu}\left(x\right)\psi+\mu\,\vec{\alpha}\cdot\vec{E}\,\psi-\mu\,\vec{\Sigma}\cdot\vec{B}\,\psi=m\psi.
\label{2.4}
\end{eqnarray}

In an inertial frame, the line element of the Minkowski spacetime background can be written as $ds^{2}=-d\mathcal{T}^{2}+d\mathcal{R}^{2}+\mathcal{R}^{2}d\Phi^{2}+d\mathcal{Z}^{2}$. In this inertial frame, we consider an electric charge density $\lambda$ concentrated on the $z$ axis. This distribution of charges creates a radial electric field in the rest frame of the observers given by $E^{\mathcal{R}}_{\mathrm{rf}}=E^{1}=\frac{\lambda\mathcal{R}}{2}$. Our interest in this work is to study the quantum dynamics of the neutral particle in a noninertial frame. Therefore, we make the coordinate transformation $\mathcal{T}=t;\,\,\,\,\mathcal{R}=\rho;\,\,\,\,\Phi=\varphi+\omega\,t;\,\,\,\,\mathcal{Z}=z$, where $\omega$ is the constant angular velocity of the rotating frame. With this transformation, the line element can be written as 
\begin{eqnarray}
ds^{2}=-\left(1-\omega^{2}\rho^{2}\right)\,dt^{2}+2\omega\rho^{2}d\varphi\,dt+d\rho^{2}+\rho^{2}d\varphi^{2}+dz^{2}.
%=-dt^{2}+d\rho^{2}+\eta^{2}\rho^{2}\left(d\varphi+\omega\,dt\right)^{2}+dz^{2}.
\label{1.5}
\end{eqnarray}

Note that this line element is valid for values of the radial coordinate inside the range $0<\rho<1/\omega$ \cite{landau3}. This means that, for all values of the radial coordinate $\rho\geq1/\omega$, the Dirac particle is placed outside of the light-cone because the velocity of the Dirac particle is greater than the velocity of the light \cite{landau3}. Hence, the line element (\ref{1.5}) is not well-defined for $\rho\geq1/\omega$. The meaning of this restriction for the values of the radial coordinate is that the wave function of a Dirac neutral particle cannot be defined everywhere, but only inside the range $0<\rho<1/\omega$. The interesting point of this restriction on the radial coordinate imposed by noninertial effects is that it gives rise to a hard-wall confining potential which imposes that the wave function $\psi\left(x\right)$ must vanish at $\rho\rightarrow 1/\omega$. We will see later that we can compare this confinement of a Dirac neutral particle with quantum dot models described by a hard-wall confining potential \cite{dot2}.

We intend to work on a well-known reference frame called the Fermi-Walker reference frame \cite{misner}. The Fermi-Walker reference frame can be built through the local frame of the observers where the spatial components of the noncoordinate basis $\hat{\theta}^{i}$, $i=1,2,3$, do not rotate and the component $\hat{\theta}^{0}$ forms a rest frame at each instant, that is, $\hat{\theta}^{0}=e^{0}_{\,\,\,t}\left(x\right)\,dt$. This frame allows us to observe noninertial effects due to the action of external forces without any effects due to arbitrary rotations of the local spatial axes \cite{misner}. In this way, we write the tetrads in the form: 
\begin{eqnarray}
%\hat{\theta}^{0}=dt;\,\,\,\hat{\theta}^{1}=d\rho;\,\,\,\hat{\theta}^{2}=\omega\rho\,dt+\rho\,d\varphi;\,\,\,\hat{\theta}^{3}=dz.
e^{a}_{\,\,\,\mu}\left(x\right)=\left(
\begin{array}{cccc}
1 & 0 & 0 & 0 \\
0 & 1 & 0 & 0 \\
\omega\rho & 0 & \rho & 0\\
0 & 0 & 0 & 1\\	
\end{array}\right);\,\,\,\,\,\,\,\,\,
e^{\mu}_{\,\,\,a}\left(x\right)=\left(
\begin{array}{cccc}
1 & 0 & 0 & 0 \\
0 & 1 & 0 & 0 \\
-\omega & 0 & \frac{1}{\rho} & 0\\
0 & 0 & 0 & 1\\	
\end{array}\right).
\label{1.8}
\end{eqnarray}

After the choice of the local reference frame of the observers, we need to obtain the components of the spinorial connection $\Gamma_{\mu}\left(x\right)$. First, we can obtain the connection 1-form $\omega^{a}_{\,\,\,b}=\omega_{\mu\,\,\,\,b}^{\,\,\,a}\left(x\right)\,dx^{\mu}$ via the Maurer-Cartan structure equations \cite{naka}. In the absence of the torsion field, the Maurer-Cartan structure equations can be write as $d\hat{\theta}^{a}+\omega^{a}_{\,\,\,b}\wedge\hat{\theta}^{b}=0$, where the operator $d$ is the exterior derivative and the symbol $\wedge$ means the wedge product. By solving the Maurer-Cartan structure equations, we have that the non-null components of the connection 1-form are $\omega_{t\,\,\,2}^{\,\,\,1}\left(x\right)=-\omega_{t\,\,\,1}^{\,\,\,2}\left(x\right)=-\omega$ and $\omega_{\varphi\,\,\,2}^{\,\,\,1}\left(x\right)=-\omega_{\varphi\,\,\,1}^{\,\,\,2}\left(x\right)=-1$. Taking these expressions for the components of the connection 1-form, we can calculate the spinorial connection $\Gamma_{\mu}\left(x\right)$ and obtain $\gamma^{\mu}\,\Gamma_{\mu}\left(x\right)=\frac{\gamma^{1}}{2\rho}$. Note that, if we consider Cartesian coordinates, the spinorial connection $\Gamma_{\mu}\left(x\right)$ vanishes \cite{bf10}. 

Our next step is to obtain the field configuration induced by the noninertial effects of the Fermi-Walker reference frame. As we have seen above, in an inertial frame, the electric field is given by $E^{1}=E^{\mathcal{R}}_{\mathrm{rf}}=\frac{\lambda\mathcal{R}}{2}$. In the Fermi-Walker reference frame (\ref{1.8}), the fields are given by the relation $F^{\mu\nu}\left(x\right)=e^{\mu}_{\,\,\,a}\left(x\right)\,e^{\nu}_{\,\,\,b}\left(x\right)\,F^{ab}\left(x\right)$ \cite{rt2,rt3,rt4}. Thus, by changing $\mathcal{R}\rightarrow\rho$, the non-null components of the electric and the magnetic fields when the local reference frames of the observers are Fermi-Walker transported are \cite{b4}
\begin{eqnarray}
E^{\rho}=\frac{\lambda\rho}{2};\,\,\,\,\,\,\,B^{z}=\frac{\omega\lambda}{2}\,\rho^{2}.
\label{1.12}
\end{eqnarray}

Now, we are able to obtain the Dirac equation for a neutral particle with a permanent magnetic dipole moment in a noninertial reference frame. Taking into account the local reference frame of the observers (\ref{1.8}) and the field configuration (\ref{1.12}), the Dirac equation (\ref{2.4}) becomes
\begin{eqnarray}
i\frac{\partial\psi}{\partial t}=m\hat{\beta}\psi+i\omega\frac{\partial\psi}{\partial\varphi}-i\hat{\alpha}^{1}\left(\frac{\partial}{\partial\rho}+\frac{1}{2\rho}\right)\psi-i\frac{\hat{\alpha}^{2}}{\rho}\frac{\partial\psi}{\partial\varphi}-i\hat{\alpha}^{3}\frac{\partial\psi}{\partial z}-i\frac{\mu\lambda}{2}\,\rho\,\hat{\beta}\,\hat{\alpha}^{1}\psi-\frac{\mu\lambda\omega}{2}\,\rho^{2}\,\hat{\beta}\,\Sigma^{3}\,\psi.
\label{2.6}
\end{eqnarray}

Note that the Dirac Hamiltonian given in the right-hand-side of (\ref{2.6}) commutes with the operators $\hat{J}_{z}=-i\frac{\partial}{\partial\varphi}$ \cite{schu} and $\hat{p}_{z}=-i\frac{\partial}{\partial z}$. Thus, in order to solve the Dirac equation (\ref{2.6}), we write the solution in the following form: $\psi=e^{-i\mathcal{E}t}\,e^{i\left(l+\frac{1}{2}\right)\varphi}\,e^{ikz}\,\left(\phi\left(\rho\right)\,\,\,\chi\left(\rho\right)\right)^{T}$, with $l=0,\pm1,\pm2,\ldots$, $k$ is a constant, $\phi=\left(\phi_{+}\,\,\,\phi_{-}\right)^{T}$ and $\chi=\left(\chi_{+}\,\,\,\chi_{-}\right)^{T}$ being two-component spinors, and where $\sigma^{3}\phi_{+}=\phi_{+}$, $\sigma^{3}\phi_{-}=-\phi_{-}$ (the same for $\chi_{\pm}$). Substituting $\psi$ into the Dirac equation (\ref{2.6}), we obtain two coupled equations of $\phi$ and $\chi$. The first coupled equation is
\begin{eqnarray}
\left[\mathcal{E}-m+\omega\left(l+\frac{1}{2}\right)-\frac{\mu\lambda\omega}{2}\,\rho^{2}\,\sigma^{3}\right]\phi=\left[-i\,\sigma^{1}\frac{\partial}{\partial\rho}-\frac{i\,\sigma^{1}}{2\rho}-i\frac{\mu\lambda}{2}\,\rho\,\sigma^{1}+\frac{\left(l+\frac{1}{2}\right)}{\rho}\,\sigma^{2}+k\,\sigma^{3}\right]\chi,
\label{2.8}
\end{eqnarray}
while the second coupled equation is
\begin{eqnarray}
\left[\mathcal{E}+m+\omega\left(l+\frac{1}{2}\right)+\frac{\mu\lambda\omega}{2}\,\rho^{2}\,\sigma^{3}\right]\chi=\left[-i\,\sigma^{1}\frac{\partial}{\partial\rho}-\frac{i\,\sigma^{1}}{2\rho}+i\frac{\mu\lambda}{2}\,\rho\,\sigma^{1}+\frac{\left(l+\frac{1}{2}\right)}{\rho}\,\sigma^{2}+k\,\sigma^{3}\right]\phi.
\label{2.9}
\end{eqnarray}

By eliminating $\chi$ from the equation (\ref{2.9}) and considering the magnetic dipole moment of the neutral particle parallel to the $z$ axis, we can neglect the terms proportional to $m^{-2}$ and $\omega^{2}\rho^{4}$ (here, we have assumed that the velocity of rotation is small when compared to the velocity of light), and obtain two non-coupled equations for $\phi_{+}$ and $\phi_{-}$. In order to write these two non-coupled equations in a compact form, we label them as $\phi_{s}$, where $s=\pm1$ and $\sigma^{3}\phi_{s}=\pm\phi_{s}=s\phi_{s}$. Since there is no torque on the dipole moment we can take $k=0$ \cite{bf10}, and obtain the following radial equation:
\begin{eqnarray}
\frac{d^{2}\phi_{s}}{d\rho^{2}}+\frac{1}{\rho}\frac{d\phi_{s}}{d\rho}-\frac{\gamma^{2}_{s}}{\rho^{2}}\,\phi_{s}-\frac{\mu^{2}\lambda^{2}\delta_{s}^{2}}{4}\,\rho^{2}\,\phi_{s}+\beta_{s}\,\phi_{s}=0,
\label{2.10}
\end{eqnarray}
where we have defined the parameters in the radial equation (\ref{2.10}):
\begin{eqnarray}
\zeta_{s}&=&l+\frac{1}{2}\left(1-s\right);\nonumber\\
\delta_{s}^{2}&=&1\pm\,\frac{4mw}{\mu\lambda};\label{2.11}\\
\beta_{s}&=&\left[\mathcal{E}+\omega\left(l+\frac{1}{2}\right)\right]^{2}-m^{2}-\,\mu\lambda\left[s\,\zeta_{s}+1\right].\nonumber
\end{eqnarray}

Let us make a change of variable given by $\xi=\frac{\mu\lambda\delta_{s}}{2}\,\rho^{2}$, and rewrite the radial equation (\ref{2.10}) in the form
\begin{eqnarray}
\xi\,\phi_{s}''+\phi_{s}'-\frac{\zeta^{2}_{s}}{4\xi}\,\phi_{s}-\frac{\xi}{4}\,\phi_{s}+\frac{\beta_{s}}{2\mu\lambda\delta_{s}}\,\phi_{s}=0.
\label{2.12}
\end{eqnarray}
In order to have the radial wave function being regular at the origin, we write the solution of equation (\ref{2.12}) in the form: $\phi_{s}\left(\xi\right)=e^{-\frac{\xi}{2}}\,\xi^{\frac{\left|\zeta_{s}\right|}{2}}\,F_{s}\left(\xi\right)$.
%\begin{eqnarray}
%\phi_{s}\left(\xi\right)=e^{-\frac{\xi}{2}}\,\xi^{\frac{\left|\zeta_{s}\right|}{2}}\,F_{s}\left(\xi\right).
%\label{2.13}
%\end{eqnarray}
Thus, substituting this solution into (\ref{2.12}), we obtain a second order differential equation 
\begin{eqnarray}
\xi\,F_{s}''+\left[\left|\zeta_{s}\right|+1-\xi\right]\,F_{s}'+\left[\frac{\beta_{s}}{2\mu\lambda\delta_{s}}-\frac{\left|\zeta_{s}\right|}{2}-\frac{1}{2}\right]\,F_{s}=0,
\label{2.14}
\end{eqnarray}
which is the Kummer equation or the confluent hypergeometric differential equation \cite{abra}. Since we wish the wave function regular at the origin, we consider only one of the solutions of the Kummer equation (\ref{2.14}) given by $F_{s}\left(\xi\right)=\,_{1}F_{1}\left[\frac{\left|\zeta_{s}\right|}{2}+\frac{1}{2}-\frac{\beta_{s}}{2\mu\lambda\delta_{s}},\left|\zeta_{s}\right|+1,\xi\right]$, which is called the Kummer function of first kind \cite{abra}. Therefore, the radial eigenfunctions are $\phi_{s}\left(\rho\right)=\left(\frac{\mu\lambda\delta_{s}}{2}\right)^{\frac{\left|\gamma_{s}\right|}{2}}\,e^{\frac{\mu\lambda\delta_{s}}{2}\rho^{2}}\,\rho^{\left|\gamma_{s}\right|}\,_{1}F_{1}\left[\frac{\left|\zeta_{s}\right|}{2}+\frac{1}{2}-\frac{\beta_{s}}{2\mu\lambda\delta_{s}},\left|\zeta_{s}\right|+1,\frac{\mu\lambda\delta_{s}}{2}\,\rho^{2}\right]$. By solving the coupled equations (\ref{2.8}) and (\ref{2.9}), we can obtain the solutions for the Dirac equation (\ref{2.6}). Thus, substituting the solution $\phi=\phi_{s}$ into Eq. (\ref{2.9}), we obtain the solutions for the two-spinor $\chi$. Hence, the positive-energy solutions of the Dirac equation (\ref{2.6}) corresponding to the parallel component of the Dirac spinor to the $z$ axis are 
\begin{eqnarray}
\psi_{+}&=&f_{+}\,F\left[\frac{\left|\zeta_{s}\right|}{2}+\frac{1}{2}-\frac{\beta_{s}}{2\mu\lambda\delta_{s}},\left|l\right|+1,\frac{\mu\lambda\delta_{+}}{2}\,\rho^{2}\right]\left(
\begin{array}{c}
1 \\
0\\
%\frac{k}{\left[\mathcal{E}+m+\omega\left(l+\frac{1}{2}\right)-\frac{\mu\lambda\omega}{2}\,\rho^{2}\right]}\\
0\\
\frac{i\left[\frac{\mu\lambda}{2}\left(1+\delta_{+}\right)\rho-\frac{\left|l\right|}{\rho}+\frac{l}{\rho}\right]}{\left[\mathcal{E}+m+\omega\left(l+\frac{1}{2}\right)-\frac{\mu\lambda\omega}{2}\,\rho^{2}\right]}\\	
\end{array}\right)\nonumber\\
[-3mm]\label{2.17}\\[-3mm]
&-&\frac{i\,f_{+}\left[\frac{\left|\zeta_{s}\right|}{2}+\frac{1}{2}-\frac{\beta_{s}}{2\mu\lambda\delta_{s}}\right]}{\left[\mathcal{E}+m+\omega\left(l+\frac{1}{2}\right)-\frac{\mu\lambda\omega}{2}\,\rho^{2}\right]}\,F\left[\frac{\left|\zeta_{s}\right|}{2}-\frac{\beta_{s}}{2\mu\lambda\delta_{s}}+\frac{3}{2},\left|l\right|+2,\frac{\mu\lambda\delta_{+}}{2}\,\rho^{2}\right]\,\left(
\begin{array}{c}
0\\
0\\
0\\
\frac{\mu\lambda \delta_{+}}{\left(\left|l\right|+1\right)}\,\rho\\	
\end{array}\right),\nonumber
\end{eqnarray}
while the antiparallel component of the Dirac spinor to the $z$ axis are
\begin{eqnarray}
\psi_{-}&=&f_{-}\,F\left[\frac{\left|\zeta_{s}\right|}{2}+\frac{1}{2}-\frac{\beta_{s}}{2\mu\lambda\delta_{s}},\left|l+1\right|+1,\frac{\mu\lambda\delta_{-}}{2}\,\rho^{2}\right]\left(
\begin{array}{c}
0 \\
1\\
\frac{i\left[\frac{\mu\lambda}{2}\left(1+\delta_{-}\right)\rho-\frac{\left|l+1\right|}{\rho}-\frac{l+1}{\rho}\right]}{\left[\mathcal{E}+m+\omega\left(l+\frac{1}{2}\right)+\frac{\mu\lambda\omega}{2}\rho^{2}\right]}\\	
%-\frac{k}{\left[\mathcal{E}+m+\omega\left(l+\frac{1}{2}\right)+\frac{\mu\lambda\omega}{2}\rho^{2}\right]}\\
0\\
\end{array}\right)\nonumber\\
[-3mm]\label{2.18}\\[-3mm]
&-&\frac{i\,f_{-}\left[\frac{\left|\zeta_{s}\right|}{2}+\frac{1}{2}-\frac{\beta_{s}}{2\mu\lambda\delta_{s}}\right]}{\left[\mathcal{E}+m+\omega\left(l+\frac{1}{2}\right)+\frac{\mu\lambda\omega}{2}\rho^{2}\right]}\,F\left[\frac{\left|\zeta_{s}\right|}{2}-\frac{\beta_{s}}{2\mu\lambda\delta_{s}}+\frac{3}{2},\left|l+1\right|+2,\frac{\mu\lambda\delta_{-}}{2}\rho^{2}\right]\,\left(
\begin{array}{c}
0\\
0\\
\frac{\mu\lambda\delta_{-}}{\left(\left|l+1\right|+1\right)}\,\rho\\	
0\\
\end{array}\right).\nonumber
\end{eqnarray}
Note that we have defined the parameters $f_{\pm}=C\,e^{-i\mathcal{E}t}\,e^{i\left(l+\frac{1}{2}\right)\varphi}\,e^{ikz}\,\left(\frac{\mu\lambda\delta_{s}}{2}\right)^{\frac{\left|\zeta_{\pm}\right|}{2}}\,e^{-\frac{\mu\lambda\delta_{s}\,\rho^{2}}{2}}\,\rho^{\left|\zeta_{\pm}\right|}$ in equations (\ref{2.17}) and (\ref{2.18}), where $C$ is a constant.
%\begin{eqnarray}
%f_{\pm}=C\,e^{-i\mathcal{E}t}\,e^{i\left(l+\frac{1}{2}\right)\varphi}\,e^{ikz}\,\left(\frac{\mu\lambda\delta_{s}}{2}\right)^{\frac{\left|\gamma_{\pm}\right|}{2}}\,e^{-\frac{\mu\lambda\delta_{s}\,\rho^{2}}{2}}\,\rho^{\left|\gamma_{\pm}\right|}.
%\label{2.19}
%\end{eqnarray}
The spinors (\ref{2.17}) and (\ref{2.18}) correspond to the positive-energy solutions of the Dirac equation (\ref{2.6}). The negative solutions of the Dirac equation (\ref{2.6}) can be obtained by using the same procedure above \cite{b4}.

Recently \cite{b4} we have discussed, by imposing the condition where the confluent hypergeometric series becomes a polynomial of degree $n$ and by assuming that $\mu\lambda\ll\omega$, that we can consider the wave function of the Dirac neutral particle is normalized in the range $0<\rho<1/\omega$ because the amplitude of probability of finding the Dirac neutral particle in the region $\rho\geq1/\omega$ is quite small. From this assumption, we have obtained the energy levels corresponding to having a Dirac neutral particle confined to a parabolic potential. Moreover, by taking the nonrelativistic limit of the energy levels, we have shown that the dispersion of the subbands is non-parabolic in a similar way of the Tan-Inkson model for a quantum dot \cite{tan}. 

In the present contribution, we do not assume the condition $\mu\lambda\ll\omega$, thus, we cannot impose the condition where the confluent hypergeometric series becomes a polynomial because the wave function cannot be considered as being normalized inside the range $0<\rho<1/\omega$ defined by the line element (\ref{1.5}). In this way, in order to obtain a normalized wave function inside the physical region of the spacetime $0\,<\,\rho\,<\,\frac{1}{\omega}$, we first impose that the radial wave function vanishes at $\rho\rightarrow1/\omega$. Note that this boundary condition arises from noninertial effects and corresponds to having the geometry of the manifold playing the role of a hard-wall confining potential. In this case, we have for a fixed radius $\rho_{0}=1/\omega$, that
\begin{eqnarray}
\phi_{s}\left(\xi_{0}=\frac{\mu\lambda\delta_{s}}{2}\,\rho_{0}^{2}\right)=0.
\label{2.21}
\end{eqnarray}

Our next step is to assume that the intensity of the electric field is given in such a way that the parameter $\mu\lambda\delta_{s}$ can be considered small. By assuming that $\mu\lambda\delta_{s}$ is quite small, and by taking a fixed value of the parameter $b=\left|\zeta_{s}\right|+1$ of the Kummer function and fixed radius $\rho_{0}=1/\omega$, then, we can consider the parameter $a=\frac{\left|\zeta_{s}\right|}{2}+\frac{1}{2}-\frac{\beta_{s}}{2\mu\lambda\delta_{s}}$ of the Kummer function being large, without loss of generality. From this assumption, we can write the Kummer function of first kind in the form \cite{abra}:  
\begin{eqnarray}
_{1}F_{1}\left(a,b,\xi_{0}=\frac{\mu\lambda\delta_{s}}{2}\,\rho_{0}^{2}\right)&\approx&\frac{\Gamma\left(b\right)}{\sqrt{\pi}}\,e^{\frac{\xi_{0}}{2}}\left(\frac{b\xi_{0}}{2}-a\xi_{0}\right)^{\frac{1-b}{2}}\,\cos\left(\sqrt{2b\xi_{0}-4a\xi_{0}}-\frac{b\pi}{2}+\frac{\pi}{4}\right),
\label{3.12}
\end{eqnarray}
where $\Gamma\left(b\right)$ is the gamma function. In this way, with $\beta_{s}$ given in (\ref{2.11}) and by using (\ref{3.12}) to rewrite  $\phi_{s}\left(\xi\right)=e^{-\frac{\xi}{2}}\,\xi^{\frac{\left|\zeta_{s}\right|}{2}}\,F_{s}\left(\xi\right)$, then, the boundary condition (\ref{2.21}) yields $\phi_{s}\left(\xi_{0}\right)\propto\sqrt{\frac{1}{\pi\beta_{s}^{\left|\zeta_{s}\right|}}}\,\cos\left(\sqrt{4\beta_{s}\xi_{0}}-\left|\zeta_{s}\right|\frac{\pi}{2}-\frac{\pi}{4}\right)=0$. Thus, we obtain the following expression for the energy levels
\begin{eqnarray}
\mathcal{E}_{n,\,l,\,s}\approx\sqrt{m^{2}+\frac{1}{\rho_{0}^{2}}\left[n\pi+\left|\zeta_{s}\right|\frac{\pi}{2}+\frac{3\pi}{4}\right]^{2}+\mu\lambda\left[s\zeta_{s}+1\right]}-\omega\left[l+\frac{1}{2}\right].
\label{3.14}
\end{eqnarray}

We have that the expression (\ref{3.14}) corresponds to the relativistic energy levels of a neutral particle with a permanent magnetic dipole moment interacting with a field configuration induced by the noninertial effects of the Fermi-Walker reference frame, when the Dirac neutral particle is subject to a hard-wall confining potential. We have seen that this hard-wall confining potential arises from noninertial effects, and it also corresponds to confining a Dirac neutral particle to a quantum dot \cite{dot2}. We can see that the relativistic energy levels (\ref{3.14}) differs from that obtained in \cite{b4} even the neutral particle is confined to moving in the range $0<\,\rho\,<\,1/\omega$. In \cite{b4}, we have considered the intensity of the electric field in such a way that $\mu\lambda\ll\omega$, then, relativistic bound state solutions analogous to having the Dirac neutral particle confined to a potential $V\left(\rho\right)=a\,\rho^{2}$ could be achieved. Here, we have imposed that the wave function vanishes at $\rho\rightarrow1/\omega$, and assumed that the intensity of the electric field is given in such a way that the parameter $\mu\lambda\delta_{s}$ can be considered small. Hence, the relativistic energy levels (\ref{3.14}) correspond to the spectrum of energy analogous to having a Dirac neutral particle confined to a quantum dot \cite{dot2}, where the geometry of the manifold plays the role of a hard-wall confining potential.

Finally, let us discuss the nonrelativistic limit of the energy levels (\ref{3.14}). The nonrelativistic limit of the energy levels is obtained by applying the Taylor expansion up to the first order term in the expression (\ref{3.14}). By doing this, the nonrelativistic energy levels become
\begin{eqnarray}
\mathcal{E}_{n,\,l}\approx m+\frac{1}{2m\rho_{0}^{2}}\left[n\pi+\left|\zeta_{s}\right|\frac{\pi}{2}+\frac{3\pi}{4}\right]^{2}+\frac{\mu\lambda}{2m}\,\left[s\zeta_{s}+1\right]-\omega\left[l+\frac{1}{2}\right],
%\left[s\,\frac{\zeta_{s}}{2}+\frac{1}{2}\right]-\omega\left[l+\frac{1}{2}\right],
\label{3.15}
\end{eqnarray}
where $m$ is the rest mass of the neutral particle, and the remaining terms of the expression (\ref{3.15}) correspond to the nonrelativistic energy levels of a neutral particle with permanent magnetic dipole moment confined to a quantum dot. Note that, in this approach, the energy levels (\ref{3.15}) are proportional to $n^{2}$ (parabolic) which differ from the results obtained in \cite{b2} whose dispersion relation is proportional to $n$ (non-parabolic) in an analogous way to the Tan-Inkson model for a quantum dot \cite{tan}. The reason for this is because we have considered another behaviour of the induced fields, and changed the boundary conditions on the wave function. We can also note that this parabolic behaviour (proportional to $n^{2}$) of the nonrelativistic spectrum of energy (\ref{3.15}) agrees with the quantum dot models described by a hard-wall confining potential given in \cite{dot2}, where the geometry of the manifold plays the role of the hard-wall confining potential. Finally, we can see in the expression (\ref{3.15}), the coupling between the quantum number $l$ and the angular velocity $\omega$ arises from noninertial effects. This coupling is called in the nonrelativistic quantum mechanics as the Page-Werner \textit{et al.} term \cite{r1,r2}.

Two interesting topics of discussion should be the arising of persistent currents and the confinement of Majorana fermions to a quantum dot in a noninertial frame. In recent decades, studies of persistent currents have been made in a quantum ring \cite{by}, two-dimensional quantum rings and quantum dots \cite{tan2} by showing the arising of persistent currents due to the dependence of the energy levels of bound states on geometric quantum phases \cite{by}. Other studies of persistent currents have been made based on the Berry phase \cite{ring2}, the Aharonov-Anandan quantum phase \cite{ring3} and the Aharonov-Casher geometric phase \cite{ring1,ring4}. Therefore, the quantum model discussed in this work can be interesting in new studies of persistent currents in Dirac-like systems \cite{graf,dirac2,dirac3,bf30} based on geometric phases induced by noninertial effects such as the Sagnac effect \cite{sag}, the Mashhoon effect \cite{r3} and the analogue of the Aharonov-Casher effect in noninertial frames \cite{bf3}. 

The second possible topic of discussion is based on the difference between a Dirac particle and a Majorana particle. As an example, by considering a neutral particle, a Dirac neutral particle is not charge self-conjugated, that is, $\psi_{c}\neq\psi$, but a Majorana neutral fermion is charge self-conjugated, $\psi_{c}=\psi$. A recent study \cite{belich3} has shown that the analogue effect of the Aharonov-Casher effect \cite{ac} for a Majorana neutral fermion cannot be obtained in a Lorentz symmetry violation background. From this difference between a Dirac neutral particle and a Majorana neutral particle, one should expect a different behaviour of a Majorana neutral particle in a noninertial system. Hence, studies of the behaviour of Majorana fermions in noninertial systems can be interesting in confinement to quantum dots \cite{bf30,dot2,b4}, and the quantum Hall effect \cite{hall1,hall2,hall7,er}.

\section{conclusions}
 
In this work, we have shown that the presence of noninertial effects gives rise to a hard-wall confining potential that acts on a Dirac neutral particle with a permanent magnetic dipole moment by confining it in analogous way to a quantum dot \cite{dot2}. However, we have seen that the relativistic bound states solutions of the Dirac equation can only be achieved depending on the behaviour of the external fields induced by the noninertial effects of the Fermi-Walker reference frame. In a previous work \cite{b4}, we have discussed a case whose intensity of the electric field satisfies the condition $\mu\lambda\ll\omega$, then, relativistic bound state solutions analogous to having the Dirac neutral particle confined to a potential $V\left(\rho\right)=a\,\rho^{2}$ could be achieved. In this work, we have discussed a new case where the condition for the intensity of the electric field $\mu\lambda\ll\omega$ is not satisfied. Here, we have shown that relativistic bound state solutions of the Dirac equation for a neutral particle with a permanent magnetic dipole moment interacting with external fields can be achieved by imposing two conditions: the first condition is that the radial wave function vanishes at $\rho\rightarrow1/\omega$, while the second condition establishes that the intensity of the electric field must be given in such a way that the parameter $\mu\lambda\delta_{s}$ can be considered small. Moreover, by comparing the nonrelativistic limit of the energy levels, we can see a parabolic behaviour (proportional to $n^{2}$) of the nonrelativistic spectrum of energy (\ref{3.15}) which agrees with the quantum dot models described by a hard-wall confining potential given in \cite{dot2}, where the geometry of the manifold plays the role of the hard-wall confining potential.

We would like to thank CNPq (Conselho Nacional de Desenvolvimento Cient\'ifico e Tecnol\'ogico - Brazil) for financial support.

\end{document}